\documentclass[conference]{IEEEtran}
\usepackage{amsmath}
\usepackage{amssymb}
\usepackage{amsfonts}
\usepackage{graphicx}
\usepackage{epsfig}
\usepackage{subfigure}
\usepackage{psfrag}
\usepackage{cite}
\usepackage{latexsym}
\usepackage{url}
\usepackage{color}
\usepackage{bm}
\PassOptionsToPackage{bookmarks={false}}{hyperref}
\IEEEoverridecommandlockouts

\begin{document}
\title{Joint Transmit and Reflective Beamforming Design for IRS-Assisted Multiuser MISO SWIPT Systems
\thanks{J. Xu is the corresponding author.}}
\author{Yizheng Tang$^1$, Ganggang Ma$^1$, Hailiang Xie$^1$, Jie Xu$^1$, and Xiao Han$^2$\\
$^1$School of Information Engineering, Guangdong University of Technology, Guangzhou, China\\
$^2$Wireless Technology Lab, 2012 Laboratories, Huawei, China\\
E-mail: eadgon\_yz\_tang@mail2.gdut.edu.cn,~gangma.gdut@gmail.com,~hailiang.gdut@gmail.com,\\~jiexu@gdut.edu.cn,~tony.hanxiao@huawei.com
}

\setlength\abovedisplayskip{2.3pt}
\setlength\belowdisplayskip{2.3pt}
\maketitle

\begin{abstract}
This paper studies an intelligent reflecting surface (IRS)-assisted multiuser multiple-input single-output (MISO) simultaneous wireless information and power transfer (SWIPT) system. In this system, a multi-antenna access point (AP) uses transmit beamforming to send both information and energy signals to a set of receivers each for information decoding (ID) or energy harvesting (EH), and a dedicatedly deployed IRS properly controls its reflecting phase shifts to form passive reflection beams for facilitating both ID and EH at receivers. Under this setup, we jointly optimize the (active) information and energy transmit beamforming at the AP together with the (passive) reflective beamforming at the IRS, to maximize the minimum power received at all EH receivers, subject to individual signal-to-interference-plus-noise ratio (SINR) constraints at ID receivers, and the maximum transmit power constraint at the AP. Although the formulated SINR-constrained min-energy maximization problem is highly non-convex, we present an efficient algorithm to obtain a high-quality solution by using the techniques of alternating optimization and semi-definite relaxation (SDR). Numerical results show that the proposed IRS-assisted SWIPT system with both information and energy signals achieves significant performance gains over benchmark schemes without IRS deployed and/or without dedicated energy signals used.
\end{abstract}

\begin{IEEEkeywords}
Intelligent reflecting surface (IRS), simultaneous wireless information and power transfer (SWIPT), joint active and passive beamforming design, optimization.
\end{IEEEkeywords}

\newtheorem{theorem}{\underline{Theorem}}[section]
\newtheorem{lemma}{\underline{Lemma}}[section]
\newtheorem{proposition}{\underline{Proposition}}[section]
\newtheorem{remark}{\underline{Remark}}[section]
\newcommand{\mv}[1]{\mbox{\boldmath{$ #1 $}}}

\section{Introduction}
Intelligent reflecting surface (IRS) has recently attracted a lot of research interests from both academia and industry as one candidate technology for the beyond fifth-generation (B5G) and sixth-generation (6G) cellular networks \cite{WuQingqing2019,MarcoDiRenzo2019}. IRS consists of a set of reconfigurable reflecting units, whose reflecting phase shifts can be adaptively adjusted to change the signal reflection environment for facilitating the wireless transmission. For instance, the IRS can properly reflect the radio-frequency (RF) signals sent from transmitters towards desirable directions, so as to efficiently enhance received signal power at intended receivers and suppress interference at unintended receivers. Furthermore, the IRS is passive antenna arrays with significantly reduced energy consumption. As a result, it is envisioned that the employment of IRS can efficiently enhance the spectrum and energy efficiency of wireless communication networks in a cost-effective manner.

In the literature, there have been several prior works investigating the joint design of active signal transmission at transmitters and passive beamforming at IRS to enhance the communication performance in IRS-assisted wireless systems (see, e.g., \cite{Wuqingqing2018,Huang2019,QQTWC,Yu2019}). For instance, \cite{Wuqingqing2018,Yu2019} considered an IRS-assited multiple-input single-output (MISO) point-to-point communication system, in which one multi-antenna transmitter communicates with one single-antenna receiver with the aid of IRS. The authors jointly optimized the transmit beamforming at the transmitter and the reflective beamforming at the IRS to maximize the received signal-to-noise ratio (SNR) at the receiver. Furthermore, \cite{QQTWC,Huang2019} studied the IRS-assited MISO broadcast system with more than one receiver. \cite{QQTWC} minimized the transmit power at the transmitter while ensuring the individual signal-to-interference-plus-noise ratio (SINR) constraints at receivers; while \cite{Huang2019} maximized the bits-per-Joule energy efficiency of the system, subject to the individual rate requirements at each receiver and the transmit power constraint at the transmitter. In addition, the joint transmission and reflection design has also been investigated in other setups such as orthogonal frequency division multiplexing (OFDM) \cite{Yang2019} and non-orthogonal multiple access (NOMA) systems  \cite{noma}, multi-cell multiuser  networks  \cite{multicell}, and secrecy communication systems \cite{secure}.

Besides sole wireless communications, simultaneous wireless information and power transfer (SWIPT) has attracted growing research interests in recent years as a viable  technique to provide simultaneous data and energy access for massive low-power devices in future Internet-of-things (IoT) networks \cite{zengyong2017,Clerckx2019}. In order to make SWIPT a reality, how to enhance the energy transmission efficiency and range, and how to efficiently balance the performance conflicts between energy harvesting (EH) and information decoding (ID) are challenging tasks to be tackled. To deal with these issues, various techniques such as adaptive power control, multi-antenna beamforming, channel estimation have been proposed (see, e.g., \cite{Clerckx2019} and the references therein). For instance, \cite{Xujie2014} proposed to use dedicated energy beamformers together with conventional information beamformers for enhancing the SWIPT performance.

Motivated by the success of IRS in wireless communications, it is expected that IRS can be a promising new solution to address the above technical issues in SWIPT systems. Nevertheless, due to the involvement of both energy and information transmission, how to design the joint transmission and reflection design to optimally balance between the EH and ID performance is becoming a new problem that remains not well addressed in the literature yet. This problem is quite challenging, especially when both energy and information beamformers are employed at transmitters. This thus motivates our investigation in this work.
\begin{figure}
\centering
 \epsfxsize=1\linewidth
    \includegraphics[width=8cm]{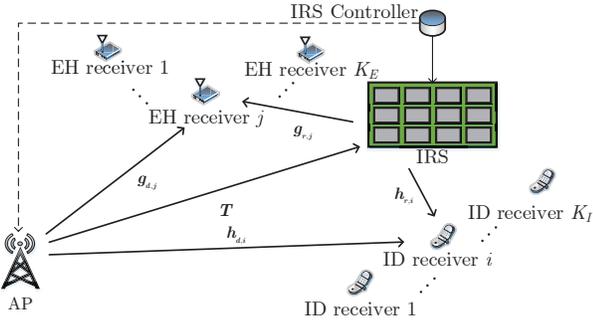}
\caption{Illustration of the multiuser MISO SWIPT system assisted by IRS.} \label{fig:1}
\vspace{-1.5em}
\end{figure}

In this paper, we consider an IRS-assisted multiuser MISO SWIPT system as shown in Fig. \ref{fig:1}, in which a multi-antenna access point (AP) uses transmit beamforming to send both information and energy signals to a set of receivers each for ID or EH, and a dedicatedly deployed IRS adaptively adjusts its reflecting phase shifts to form passive beams for facilitating both ID and EH at receivers. It is assumed that the energy signals are generated at the AP as pseudo-random sequences that are {\it a-priori} known, such that each ID receiver can perfectly cancel the resultant interference \cite{Xujie2014}. Under this setup, we jointly optimize the active transmit information and energy beamforming vectors at the AP and the passive reflective beamforming vectors at the IRS, in order to maximize the minimum received power at all EH receivers, subject to the individual SINR constraints at ID receivers and the maximum transmit power constraint at the AP. Although the formulated SINR-constrained min-energy maximization problem is highly non-convex, we propose an efficient algorithm to obtain a high-quality solution based on the alternating optimization and semidefinite relaxation (SDR) techniques. It is observed that at the obtained solution, multiple energy beams are generally required to maximize the min-energy at EH receivers. Numerical results show that our proposed IRS-assisted SWIPT system with both information and energy signals achieves significant performance gains over benchmark schemes without IRS deployed and/or without dedicated energy signals used.

It is worth noting that this paper is significantly different from a prior work \cite{Wuqingqing2019SWIPT} that also investigated the joint active and passive beamforming in IRS-assisted SWIPT systems. While \cite{Wuqingqing2019SWIPT} aimed to maximize the weighted sum-energy at EH receivers with at most one single energy beam required, this paper considers the min-energy maximization, for which multiple energy beams are generally required to ensure the energy fairness among EH receivers. Furthermore, in \cite{Wuqingqing2019SWIPT} the ID receivers were assumed to not have the capability of canceling the interference from energy signals, while this work considers a different type of ID receivers that have this capability to achieve enhanced SWIPT performance \cite{Xujie2014}.

\section{System Model}\label{sec:system}
As shown in Fig. \ref{fig:1}, we consider an IRS-assisted multiuser MISO SWIPT system, where a multi-antenna AP simultaneously sends information and energy signals to $K_I$ ID receivers and $K_E$ EH receivers, with the assistance of an IRS. Suppose that the AP is equipped with $M>1$ antennas, each ID or EH receiver is deployed with a single antenna, and the IRS has $N$ reflecting units. We define $\mathcal M \triangleq\{1,\ldots,M\}$ as the set of AP's transmit antennas, $\mathcal {K_I}\triangleq\{1,\ldots,K_I\}$ the set of ID receivers, $\mathcal {K_E}\triangleq\{1,\ldots,K_E\}$ the set of EH receivers, and $\mathcal N\triangleq\{1,\ldots,N\}$ the set of IRSs' reflecting units, respectively. Let $\mv h_{d,i}\in\mathbb{C}^{M\times 1}$ and $\mv g_{d,j}\in\mathbb{C}^{M\times 1}$ denote the channel vectors from the AP to ID receiver $i \in \mathcal{K_I}$ and EH receiver $j \in\mathcal{K_E}$, and $\mv h_{r,i}\in\mathbb{C}^{N\times 1}$ and $\mv g_{r,j}\in\mathbb{C}^{N\times 1}$ denote those from the IRS to ID receiver $i$ and EH receiver $j$, respectively. Furthermore, let $\mv T\in\mathbb C^{N \times M}$ denote the channel matrix from the AP to the IRS. It is assumed that to facilitate the joint active and passive beamforming design, the AP perfectly knows the global channel state information (CSI), similarly as in prior works \cite{Yu2019,QQTWC,Huang2019,Yang2019,noma,multicell}.

First, we consider the transmit information and energy beamforming at the AP. Let $s_i$ denote the information-carrying signal for ID receiver $i \in \mathcal{K_I}$, and $\mv w_i\in\mathbb{C}^{M\times 1}$ the corresponding information beamforming vector. Here, $s_i$ is assumed to be a circularly symmetric complex Gaussian (CSCG) random variable with zero mean and unit variance, i.e., $s_i \sim \mathcal{CN}(0, 1)$. Also, let $\mv s_E\in \mathbb{C}^{M \times 1}$ denote the dedicated energy-bearing signal, which is generated at the AP as pseudo-random sequences with mean zero and covariance ${\mv S_E} \triangleq {\mathbb E \left(\mv s_E \mv s_E^H\right)}\succeq \mv 0$. Here, the superscript $H$ denotes the conjugate transpose of a vector/matrix, $\mv A\succeq \mv 0$ means that matrix $\mv A$ is positive semi-definite, and $\mathbb{E}(\cdot)$ denotes the statistical expectation. Notice that $\mv S_E$ can be of high rank (${\rm{rank}}(\mv S_E) > 1$ may hold) in general, which corresponds to the case that a number of ${\rm{rank}}(\mv S_E)$ energy beams are delivered. By combining the information and energy signals, the transmitted signal at the AP is $\sum_{i\in \mathcal{K_I}} \mv w_i s_i + \mv s_E$. In this case, the total transmit power at the AP is expressed as $\mathbb{E}\left(\left|\sum_{i\in \mathcal{K_I}} \mv w_i s_i + \mv s_E\right|^2\right) =  \sum_{i\in\mathcal{K_I}}\|\mv w_i\|^2 + {\rm{tr}}(\mv S_E)$, where $\|\cdot\|$ denotes the Euclidean norm of a vector. Suppose that  $P$ denotes the maximum transmit power at the AP. Then we have $\sum_{i\in\mathcal{K_I}}\|\mv w_i\|^2 + {\rm{tr}}(\mv S_E)\le P$.

Next, we consider the passive reflective beamforming at the IRS. Different from the active transmit transmission at the AP, the IRS can only passively reflect the received signals by controlling the $N$ reflecting phase shifts. Let $\theta_n \in [0,2\pi)$ denote the phase shift at reflecting unit $n\in \mathcal N$. Accordingly, define $\mv \theta=[\theta_1,\ldots,\theta_N]$ and  $\mv \Theta={\rm{diag}}(e^{j\theta_1},\ldots,e^{j\theta_n},\ldots,e^{j\theta_N})$ as the reflecting phase-shifting vector and the corresponding passive beamforming matrix, respectively. Here, ${\rm {diag}}(a_1, ..., a_N)$ denotes a diagonal matrix with diagonal elements $a_1,... , a_N$, and $j = \sqrt{-1}$.

By combining the transmitted signal from the AP and the reflected signal from the IRS, the received signal at ID receiver $i\in \mathcal{K_I}$ is expressed as
\begin{align}
&{y_i}=(\mv h_{r,i}^H\mv \Theta\mv T+\mv h_{d,i}^H)\bigg(\sum\limits_{i\in\mathcal{K_I}}{\mv w_i s_i}+\mv s_E\bigg)+n_i,\label{equa:EH:power}
\end{align}
where $n_i \sim \mathcal{CN}(0,\sigma^2_i)$ denotes the additive white Gaussian noise (AWGN) at ID receiver $i$ with noise power $\sigma^2_i$. As $\mv s_E$ is generated at the AP as pseudo-random sequences that are {\it a-priori} known, we assume that each ID receiver $i$ is able to perfectly cancel the interference from energy signals (i.e., $(\mv h_{r,i}^H\mv \Theta\mv T+\mv h_{d,i}^H)\mv s_E$) prior to decoding the desirable signal $s_i$ \cite{Xujie2014}. After such interference cancellation, the received SINR at ID receiver $i\in\mathcal{K_I}$ is given as
\begin{align}
\gamma_i(\{\mv w_i\},\mv\theta)\!=\!\frac{{|({\mv h}_{r,i}^H\mv{\Theta}\mv T+{\mv h}_{d,i}^H){\mv w}_i|}^2}{\sum\limits_{k\ne i,k\in \mathcal{K_I}}{{|({\mv h}_{r,i}^H\mv{\Theta}\mv T+{\mv h}_{d,i}^H){\mv w}_k|}^2}+\sigma^2_i}.\label{equa:DI:SINR}
\end{align}

On the other hand, each EH receiver can harvest RF energy from both information and energy signals. Accordingly, the received RF power at EH receiver $j\in\mathcal{K_E}$ is
\begin{align}
&Q_j(\{\mv w_i\},\mv S_E,\mv\theta)\nonumber\\
=~&\mathbb{E}\left(\left|(\mv g_{r,j}^H\mv\Theta\mv T+\mv g_{d,j}^H)\bigg(\sum\limits_{i\in\mathcal{K_I}}{\mv w_i s_i}+\mv s_E\bigg )\right|^2\right)\label{equa:EH:Q}\\
=~&{(\mv g_{r,j}^H\mv \Theta\mv T+\mv g_{d,j}^H)\mv S_E(\mv T^H\mv \Theta^H\mv g_{r,j}+\mv g_{d,j})}\nonumber\\
&+\sum\limits_{i\in\mathcal{K_I}}{|(\mv g_{r,j}^H\mv{\Theta}\mv T+\mv g_{d,j}^H)\mv w_i|^2}.\label{equa:EH:energy}
\end{align}

In order to balance the performance tradeoff between EH and ID receivers, our objective is to maximize the minimum received RF power among all the EH receivers{\footnote{In practice, although the RF-to-direct current (DC) conversion process at each EH receiver is highly nonlinear, the harvested DC power is generally a monotonically non-decreasing function with respect to the received RF power (see, e.g., \cite{Clerckx2019}). Therefore, maximizing the minimum RF power at EH receivers (see problem (P1)) is actually equivalent to maximizing their minimum harvested DC power.}}, subject to the minimum SINR requirement at each ID receiver and the maximum transmit power constraint at the AP. The decision variables include both the active information beamforming vectors $\{\mv w_i\}$ and energy covariance matrix $\mv S_E$ at the AP, and the passive beamforming vector (i.e., the reflective phase-shifting vector $\mv\theta$) at the IRS. The SINR-constrained min-energy maximization problem is thus formulated as
\begin{align}
{\mathtt{(P1)}}:&\max_{\{{\bm w}_i\},\bm S_E,\bm\theta}\min_{\ j\in\mathcal{K_E}}{Q_j(\{\mv w_i\},\mv S_E,\mv\theta)}\label{equa:Optproblem:1}\\
&~~~~~{\mathrm{s.t.}}~~~\gamma_i(\{\mv w_i\},\ \mv\theta)\ge\Gamma_i\ \forall i\in\mathcal{K_I}\label{equa:SINR:constraint1}\\
&~~~~~~~~~~~~\sum\limits_{i\in\mathcal{K_I}}{\|\mv w_i\|^2}+{\rm{tr}}(\mv S_E)\le P\label{equa:power:constraint1}\\
&~~~~~~~~~~~~~0 \le\theta_n \le 2\pi,\ \forall n\in\mathcal{N}\label{equa:phase:constraint1}\\
&~~~~~~~~~~~~~\mv S_E\succeq 0,\label{equal:W_E:constraint1}
\end{align}
where $\Gamma_i$ denotes the minimum SINR threshold at each ID receiver $i\in\mathcal{K_I}$. It is observed that problem $\mathtt{(P1)}$ is highly non-convex due to the coupling between the active beamforming vectors (i.e., $\{\mv w_i\}$ and $\mv S_E$) and the passive reflective phase-shifting vector $\mv \theta$. Therefore, problem $\mathtt{(P1)}$ is very challenging to be optimally solved.

\section{Proposed Solution to Problem ($\mathtt{P1}$)}\label{sec:solution}
In this section, we propose an efficient algorithm to find a high-quality solution to problem (P1). To facilitate the derivation, we first introduce an auxiliary variable $t$, and reformulate problem (P1) as the following equivalent problem:
\begin{align}
&{\mathtt{(P2)}}:\max_{\{\bm w_i\},\bm S_E,\bm\theta,t}~t\nonumber\\
&~~~~~~~~~~~~~~{\mathrm{s.t.}}~~~{Q_j(\{\mv w_i\},\mv S_E,\mv\theta)}\ge t,\ \forall j\!\in\!\mathcal{K_E}\label{equa:receivedpower:constraint2}\\
&~~~~~~~~~~~~~~~~~~~~~(\ref{equa:SINR:constraint1}),~(\ref{equa:power:constraint1}),~(\ref{equa:phase:constraint1}),\ \text{and}~(\ref{equal:W_E:constraint1}).\nonumber
\end{align}
Then, we solve problem (P2) by using the technique of alternating optimization, in which the active beamforming vectors (i.e., $\{\mv w_i\}$ and $\mv S_E$) and the passive beamforming or phase-shifting vector (i.e., $\mv\theta$) are optimized in an alternating manner, by considering the other to be given.

\subsection{Active Beamforming Optimization}\label{sec:solution:A}
First, we consider the optimization of $\{\mv w_i\}$ and $\mv S_E$ under any given $\mv \theta$. By defining the combined channel vectors $\mv h_i=\mv T^H\mv \Theta^H\mv h_{r,i}+\mv h_{d,i}$, $\forall i\in\mathcal{K_I}$, and $\mv g_j=\mv T^H\mv \Theta^H\mv g_{r,j}+\mv g_{d,j}$, $\forall j\in\mathcal{K_E}$, the active beamforming optimization problem is expressed as
\begin{align}
&{\mathtt{(P3)}}:\nonumber\\&\max_{\{\bm w_i\},\bm S_E,t}~~t\nonumber\\
&~~~~~{\mathrm{s.t.}}~\sum\limits_{i\in\mathcal{K_I}}{|\mv w_i^H \mv g_j|^2}+{\mv g_{j}^H\mv S_E\mv g_{j}}\ge t,\ \forall j\in\mathcal{K_E}\label{equa:receivedpower:constraint3}\\
&~~~~~~~~~~~\frac{|\mv w_i^H \mv h_i|^2}{\Gamma_i}-\sum\limits_{k\ne i,k\in \mathcal{K_I}}{|\mv w_k^H \mv h_i|^2}-\sigma^2_i\ge 0,\ \forall i\in\mathcal{K_I}\label{equa:SINR:constraint3}\\
&~~~~~~~~~~~(\ref{equa:power:constraint1})~\text{and}~(\ref{equal:W_E:constraint1}).\nonumber
\end{align}

Problem (P3) is a non-convex quadratically constrained quadratic program (QCQP) due to the non-convex quadratic constraints in (\ref{equa:receivedpower:constraint3}) and (\ref{equa:SINR:constraint3}). To tackle this issue, we use the SDR technique to solve (P3). Towards this end, we define $\mv W_i=\mv w_i\mv w_i^H\succeq \mv 0$ with ${\rm{rank}}(\mv W_i)\le 1, \forall i\in\mathcal{K_I}$. Then problem ({P3}) is equivalent to the following problem (P3.1).
\begin{align}
&{\mathtt{(P3.1)}}:\nonumber\\&\max_{\{\bm W_i\},\bm S_E,t}~t\nonumber\\
&{\mathrm{s.t.}}\sum\limits_{i\in\mathcal{K_I}}{{\rm{tr}}(\mv g_j\mv g_j^H\mv W_i)}+{\rm{tr}}(\mv g_j\mv g_j^H\mv S_E)\ge t,\ \forall j\in\mathcal{K_E}\label{equa:receivedpower:constraintP3.1}\\
&\frac{{\rm{tr}}(\mv h_i\mv h_i^H\mv W_i)}{\Gamma_i}-\sum\limits_{k\neq i,k\in\mathcal{K_I}}{{\rm{tr}}(\mv h_i\mv h_i^H\mv W_k)}-\sigma^2_i\ge 0,\ \forall i\in\mathcal{K_I}\label{equa:SINR:constraintP3.1}\\
&~~~~~\sum\limits_{i\in\mathcal{K_I}}{{\rm{tr}}(\mv W_i)}+{\rm{tr}}(\mv S_E)\le P\label{equa:power:constraintP3.1}\\
&~~~~~~\mv S_E\succeq 0,\ \mv W_i\succeq 0,\ \forall i\in\mathcal{K_I}\label{equa:W_i:constraintP3.1}\\
&~~~~~~{\rm {rank}}(\mv W_i) \le 1,\ \forall i\in\mathcal{K_I}\label{equa:rankone:constraintP3.1}
\end{align}

However, problem ${\mathtt{(P3.1)}}$ is still non-convex due to the non-convex rank-one constraints in (\ref{equa:rankone:constraintP3.1}). By relaxing these rank-one constraints, the SDR of problem (P3.1) is expressed as
\begin{align}
{\mathtt{(P3.2)}}:\max_{\{\bm W_i\},\bm S_E,t}&~~t\nonumber\\
{\mathrm{s.t.}}~~&(\ref{equa:receivedpower:constraintP3.1}),~(\ref{equa:SINR:constraintP3.1}),~(\ref{equa:power:constraintP3.1}),~\text{and}~(\ref{equa:W_i:constraintP3.1}).\nonumber
\end{align}
It is evident that (P3.2) is a convex semi-definite program (SDP) \cite{M.Grant_And_S.Boyd}, which can thus be efficiently solved by using standard convex optimization tools such as CVX \cite{cvx}. We denote the obtained optimal solution to problem (P3.2) as $\{\mv W^{\star}_i\}$, $\mv S^{\star}_E$, and $ t^{\star}$. Notice that if the obtained $\mv W_i^\star$'s are all of rank one, then the obtained $\{\mv W_i^{\star}\}$, $\mv S_E^\star$, and $t^\star$ are also the optimal solution to problem (P3.1). Otherwise, if any one of the obtained $\mv W^{\star}_i$'s is of high rank, then the following randomization step is needed to construct a rank-one solution to problem (P3.1).

Towards this end, we first perform the eigenvalue decomposition (EVD) over the obtained high-rank information beamformers $\mv W_i^{\star}$'s, i.e., $\mv W_i^{\star}=\mv U_{\bm W_i}\mv\Sigma_{\bm W_i}\mv U_{\bm W_i}^H, \forall i\in\mathcal{K_I}$. Accordingly, we construct the information beamforming vectors as $\bar{\mv w}_i=\mv U_{\bm W_i}\mv \Sigma_{\bm W_i}^{\frac {1}{2}}\mv r_i$ and accordingly define $\bar{\mv W}_i=\bar{\mv w}_i\bar{\mv w}_i^H, \forall i\in\mathcal{K_I}$, where $\mv r_i \sim \mathcal{CN}(\mv 0,\mv I)$ is a random CSCG vector with zero mean and covariance matrix $\mv I$. However, the newly constructed rank-one matrices $\bar{\mv W}_i$'s may not satisfy the SINR constraints in (\ref{equa:SINR:constraintP3.1}) and the power  constraint in (\ref{equa:power:constraintP3.1}), and thus may be infeasible for problem (P3.1). Inspired by \cite{solution}, we adopt an additional power optimization step to obtain a feasible and efficient solution to problem (P3.1) after the randomization process. In particular, we define $\hat{\mv W}_i = \bar{\mv W}_i/{\rm{tr}}(\bar{\mv W}_i)$ with ${\rm{rank}}(\hat{\mv W}_i)=1$ and ${\rm{tr}}(\hat{\mv W}_i)=1$, and denote $p_i$ as the transmit power at the AP for ID receiver $i \in \mathcal{K_I}$ to be optimized. Accordingly, by substituting ${\mv W}_i = p_i \hat{\mv W}_i$ and replacing $\mv S_E$ as $\mv S_E^\star$ in problem (P3.2), we have the power optimization problem as
\begin{align}
\max_{\{p_i\},t}&~~t\label{power:op}\\
{\mathrm{s.t.}}~&\sum\limits_{i\in\mathcal{K_I}}{p_i{\rm{tr}}(\mv g_j\mv g_j^H\hat{\mv W}_i)}+{\rm{tr}}(\mv g_j\mv g_j^H\mv S^{\star}_E)\ge t,\ \forall j\in\mathcal{K_E}\nonumber\\
&\frac{p_i{\rm{tr}}(\mv h_i\mv h_i^H\hat{\mv W}_i)}{\sum\limits_{k\neq i,k\in\mathcal{K_I}}{p_k{\rm{tr}}(\mv h_i\mv h_i^H\hat{\mv W}_k)}-\sigma^2_i}\ge \Gamma_i ,\ \forall i\in\mathcal{K_I}\nonumber\\
&\sum\limits_{i\in\mathcal{K_I}}{p_i}+{\rm{tr}}(\mv S^{\star}_E)\le P.\nonumber
\end{align}
It is evident that problem (\ref{power:op}) is a convex linear program (LP) \cite{M.Grant_And_S.Boyd}, which can be optimally solved by using CVX. Let $\{p_i^{\star\star}\}$ and $t^{\star\star}$ denote the optimal solution to problem (\ref{power:op}). Accordingly, we obtain the rank-one solution of $\mv W_i$ to problem (P3.1) as $\mv W^{\star\star}_i=p_i^{\star\star}\hat{\mv W}_i$, and the corresponding information beamforming vector solution to problem (P3) as $\mv w^{\star\star}_i$ with $\mv W^{\star\star}_i= \mv w^{\star\star}_i \mv w^{\star\star H}_i$, $\forall i\in \mathcal{K_I}$. By combining $\{\mv w^{\star\star H}_i\}$ together with $t^{\star\star}$ and $\mv S^{\star}$, an efficient solution to problem (P3) is finally obtained. Notice that to ensure the quality of the obtained solution to (P3), multiple randomization processes need to be implemented, and the one achieving the maximum
objective value of (P3) should be selected \cite{solution}.

\begin{remark}\label{remark:1}
It is worth emphasizing that at the obtained solution, ${\rm{rank}}(\mv S^\star_E) > 1$ may hold, such that more than one energy beams are required to maximize the min-energy transferred to EH receivers. This can be intuitively understood by considering the pure WPT case with the number of ID receivers $K_I=0$, and supposing that the number of EH receivers $K_E$ are sufficiently large. In this case, we must have $\mv S^\star_E$ to be of full rank, since otherwise, there may always exist an EH receiver with the corresponding channel vector being orthogonal to $\mv S^\star_E$, thus leading to zero received power. This phenomenon has been revealed in prior work on multiuser WPT \cite{XuWCL}. This is in sharp contrast to the weighted sum-energy maximization problem for IRS-assisted SWIPT \cite{Wuqingqing2019SWIPT}, in which only one single energy beam is required.
\end{remark}

\subsection{Passive Beamforming Optimization}\label{sec:solution:B}
In this subsection, we optimize the passive baemforming or phase-shifting vector $\mv \theta$ under any given active beamformers (i.e., $\{\mv w_i\}$ and $\mv S_E$). In this case, the passive beamforming optimization problem becomes
\begin{align}
&{\mathtt{(P4)}}:\nonumber\\
&\max_{\bm\theta,t}~~~t\nonumber\\
&~~{\mathrm{s.t.}}~{(\mv g_{r,j}^H\mv \Theta\mv T+\mv g_{d,j}^H)\mv S_E(\mv T^H\mv \Theta^H\mv g_{r,j}+\mv g_{d,j})}+\nonumber\\
&~~~~~~\sum\limits_{i\in\mathcal{K_I}}{|(\mv g_{r,j}^H\mv{\Theta}\mv T+\mv g_{d,j}^H)\mv w_i|^2} \ge t,\ \forall j\in\mathcal{K_E}\label{equa:receivedpower:constraintP4}\\
&~~~~~~\frac{|({\mv h}_{r,i}^H\mv{\Theta}\mv T+{\mv h}_{d,i}^H){\mv w_i}|^2}{\Gamma_i}\!-\!\sum\limits_{k\neq i,k\in\mathcal{K_I}}{|({\mv h}_{r,i}^H\mv{\Theta}\mv T+{\mv h}_{d,i}^H){\mv w_k}|}\nonumber\\
&~~~~~~-\sigma^2_i\ge 0,\ \forall i\in\mathcal{K_I}\label{equa:SINR:constraintP4}\\
&~~~~~~~(\ref{equa:phase:constraint1}).\nonumber
\end{align}

Note that for problem (P4), the passive beamforming vector $\mv \theta$ is embedded in the diagonal matrix $\mv \Theta$, which makes the constraints (\ref{equa:receivedpower:constraintP4}) and (\ref{equa:SINR:constraintP4}) non-convex over $\mv\theta$. Motivated by \cite{Wuqingqing2018}, we transform them into convex forms by using the following algebraic transforming. First, by supposing ${\rm rank}(\mv S_E) = r_E$, we can express $\mv S_E$ as $\mv S_E=\sum_{k=1}^{r_E} \mv v_k\mv v_k^H$ based on EVD. Accordingly, we define
\begin{align}
&\mv C_{k,i}=
\begin{bmatrix}
\mv c_{k,i}\mv c_{k,i}^H & {\mv c_{k,i}}d_{k,i}^H\\
{\mv c_{k,i}^H}d_{k,i} & 0
\end{bmatrix}
\ \forall k\in\mathcal{K_I},\ i\in\mathcal{K_I}.\label{matrix:C}\\
&\mv E_{j,i}=
\begin{bmatrix}
\mv e_{j,i}\mv e_{j,i}^H & {\mv e_{j,i}}f_{j,i}^H\\
{\mv e_{j,i}^H}f_{j,i} & 0
\end{bmatrix}
\ \forall j\in\mathcal{K_E},\ i\in\mathcal{K_I}.\label{matrix:E}\\
&\mv O_{j,k}=
\begin{bmatrix}
\mv o_{j,k}\mv o_{j}^H & {\mv o_{j,k}}q_{j,k}^H\\
{\mv o_{j,k}^H}q_{j,k} & 0
\end{bmatrix}\ \forall j\in\mathcal{K_E},\ k\in\{1,\ldots,r_E\}.\label{matrix:O}
\end{align}
Here, $\mv c_{k,i}={\rm diag}(\mv h_{r,k}^H)\mv T\mv w_i$, $d_{k,i}=\mv h_{d,k}^H\mv w_i$, $\mv e_{j,i}={\rm diag}(\mv g_{r,j}^H)\mv T\mv w_i$, $f_{j,i}=\mv g_{d,j}^H\mv w_i$, and  $\mv o_{j,k}={\rm{diag}}(\mv g_{r,j}^H)\mv T\mv v_k$, and $q_{j,k}=\mv g_{d,j}^H\mv v_k$. Next, we define $\mv\phi=[e^{-j\theta_1},\ldots,e^{-j\theta_N}, l]$ with $l^2=1$, and then we have the passive beamforming matrix as $\mv\Phi={{\mv{{\phi}}}^H}\mv{{\phi}}$ with ${\rm rank}(\mv\Phi)=1$. As a result, problem (P4) is equivalent to
\begin{align}
&{\mathtt{(P4.1)}}:\nonumber\\
&\max_{\bm \Phi,t}~~~t \nonumber\\
&~~{\mathrm{s.t.}}~\sum\limits_{i\in\mathcal{K_I}}{\rm{tr}}(\mv E_{j,i}\mv \Phi)+\sum\limits_{k=1}^{r_E}{{\rm{tr}}(\mv O_{j,k}\mv \Phi)}+\sum\limits_{i\in\mathcal{K_I}}{|f_{j,i}|^2}\nonumber\\
&~~~~~~~~+\sum\limits_{k=1}^{r_E}{|q_{j,k}|^2}\ge t,\ \forall j\in\mathcal{K_E}\label{equa:power:constraintP4.1}\\
&~~~~~~~~~{\rm{tr}}(\mv C_{i,i}\mv \Phi)+|d_{i,i}|^2\ge{\Gamma}_i\sum\limits_{k\neq i, k\in\mathcal{K_I}}{{\rm{tr}}(\mv C_{i,k}\mv \Phi)}\nonumber\\
&~~~~~~~~~+{\Gamma}_i(\sum\limits_{k\neq i,\forall k\in\mathcal{K_I}}{|d_{i,k}|^2}+\sigma^2_i),~\forall i\in\mathcal{K_I}\label{equa:SINR:contraintP4.1}\\
&~~~~~~~~~\Phi_{n,n}=1,\ \forall n\in\{1,\ldots,N+1\}\label{equa:diag=1:constraintP4.1}\\
&~~~~~~~~~\mv \Phi\succeq 0\label{equa:Phi£ºconstraintP4.1}\\
&~~~~~~~~~{\rm {rank}}(\mv\Phi)=1.\label{equa:phaserank:constratintP4.1}
\end{align}
However, problem (P4.1) is still non-convex, due to the rank-one constraint in (\ref{equa:phaserank:constratintP4.1}). By relaxing this constraint, the SDR of problem (P4.1) is given as
\begin{align}
{\mathtt{(P4.2)}}:&\max_{\bm \Phi,t}~~~~t \nonumber\\
&~~{\mathrm{s.t.}}~~~(\ref{equa:power:constraintP4.1}),~(\ref{equa:SINR:contraintP4.1}),
~(\ref{equa:diag=1:constraintP4.1}),~\text{and}~(\ref{equa:Phi£ºconstraintP4.1}).\nonumber
\end{align}
It is clear that problem (P4.2) is a convex SDP that can be optimally solved via CVX. We denote the obtained optimal solution to (P4.2) as $\mv\Phi^{*}$ and $t^{*}$.

Note that if $\mv\Phi^{*}$ is of rank one, then the solution of $\mv\Phi^{*}$ and $t^{*}$ is also optimal to problem (P4.1). In this case, by performing EVD over $\mv\Phi^{*}$, we can obtain the corresponding $\mv\phi^{*}$, and accordingly have the optimal solution to problem (P4) as $\mv\theta^{*}={\rm arg}({\mv{\phi}}^{*}/{\mv{\phi}^{*}_{N+1}})_{[1:N]}$, where ${\rm{arg}(\mv x)}$ denotes a phase vector with each element being the corresponding phase of a vector $\mv x$ and $(\cdot)_{1:N}$ extracts the first $N$ elements in a vector. Otherwise, if $\mv\Phi^{*}$ is of high rank, we need to further construct the rank-one solution as follows. Expressing the EVD of $\mv\Phi^{*}$ as $\mv\Phi^{*}=\mv U_{\mv\Phi}\mv\Sigma_{\mv\Phi}\mv U_{\mv\Phi}^H$, we then set ${\mv\phi}^*=\mv U_{\mv\Phi}\mv \Sigma_{\mv\Phi}^{\frac {1}{2}}\mv r$, where $\mv r\sim \mathcal{CN}(\mv 0,\mv I)$ is a random CSCG vector with zero mean and covariance matrix $\mv I$. Accordingly, the obtained solution to problem (P4) is $\mv\theta^{*}={\rm  arg}({\mv{\phi}}^{*}/{\mv{\phi}^{*}_{N+1}})_{[1:N]}$. Notice that the randomization needs to be implemented multiple times and the vector ${\mv{\phi}}^*$ is selected as the one achieving the maximum objective value of (P4). Therefore, the solution to problem (P4) is finally obtained.

Finally, with the solutions to problems (P3) and (P4) at hand, problem (P2) or equivalently (P1) is solved via the alternating optimization, in which problems (P3) and (P4) are solved in an alternating manner, until the convergence.

\section{Numerical Results}

In this section, we provide numerical results to validate the performance of the proposed IRS-assisted SWIPT system. In the simulation, we consider Rician fading for the wireless channel from the AP to the IRS with the Rician factor being $10$ dB, and Rayleigh fading for the wireless channels from the AP and the IRS to the receivers. We consider the path-loss model $\kappa\left(\frac{d}{d_0}\right)^{-\alpha}$, where $\kappa=-30$~dB corresponds to the path-loss at the reference distance of $d_0=1$~m. The path-loss exponent is set as $\alpha=2$ for the AP-IRS link, $\alpha=3.5$ for AP-receiver links, and $\alpha=2.5$ for IRS-receiver links. Furthermore, we set the distance from the AP to each EH receiver as $d = 3$ m, that from AP to each ID receiver as $d = 50$ m, and that from the AP to the IRS as $d = 8$ m. We also set the number of reflecting units at the IRS to be $N=40$.

\begin{figure}
\centering
 \epsfxsize=1\linewidth
    \includegraphics[width=7.5cm]{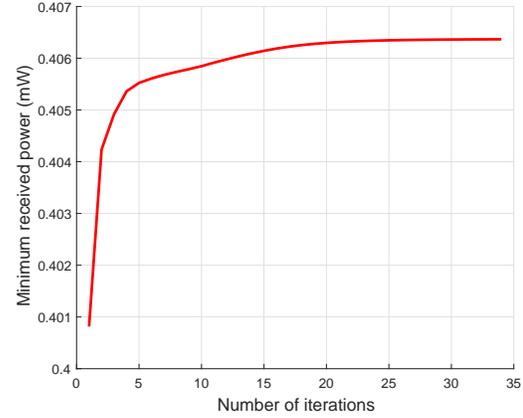}
\caption{Convergence behavior of the proposed algorithm for solving problem (P2) or (P1).} \label{fig:convergence}
\vspace{-1em}
\end{figure}
First, Fig. \ref{fig:convergence} shows the convergence behavior of our proposed alternating-optimization-based approach for solving problem (P2) or (P1). It is observed that the algorithm converges within 20-30 iterations, thus validating its effectiveness.

Next, we compare the performance of the IRS-assisted SWIPT system with both information and energy signals, versus the following benchmark schemes without IRS deployed and/or without dedicated energy signals used.

1) {\it Information beamforming only with IRS}: The AP sends information beams $\{\mv w_i\}$ to convey both information and energy, with $\mv S_E = \mv 0$. In this case, the SINR-constrained min-energy maximization problem is formulated as
\begin{align}
\max_{\{{\bm w}_i\},\bm\theta}~&\min_{\ j\in\mathcal{K_E}}~\sum\limits_{i\in\mathcal{K_I}}{|(\mv g_{r,j}^H\mv{\Theta}\mv T+\mv g_{d,j}^H)\mv w_i|^2}\label{S1:problem}\\
{\mathrm{s.t.}}~
&\sum\limits_{i\in\mathcal{K_I}}{\|\mv w_i\|^2}\le P\label{equa:power:B1}\\
& (\ref{equa:SINR:constraint1})~\text{and}~(\ref{equa:phase:constraint1}).\nonumber
\end{align}
It is observed that problem (\ref{S1:problem}) has a similar structure as (P1), and thus can be solved via a similar approach as in Section \ref{sec:solution} based on the alternating optimization and SDR.

2) {\it Conventional design without IRS}: The AP transmits both information and energy beams for SWIPT, but no IRS is deployed. The SINR-constrained min-energy maximization corresponds to problem (P1) with $\mv \Theta=\mv 0$, which can be solved by using the SDR technique together with Gaussian randomization, similarly as in Section \ref{sec:solution:A}.

3) {\it Information beamforming only without IRS}: The AP transmits only information beams for SWIPT, and no IRS is deployed. This corresponds to problem (\ref{S1:problem})
 with $\mv \Theta=\mv 0$, which can be solved similarly as in Section \ref{sec:solution:A}.

\begin{figure}
\centering
 \epsfxsize=1\linewidth
    \includegraphics[width=8cm]{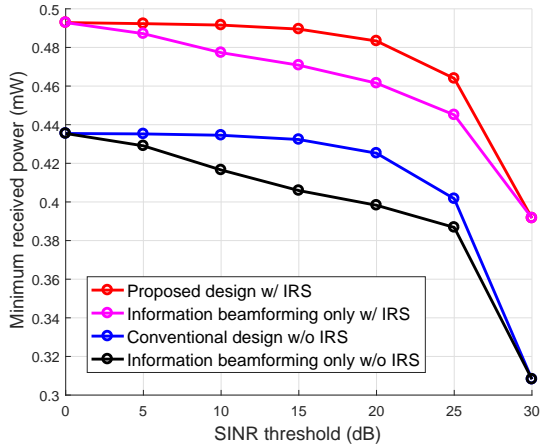}
\caption{The minimum received power at EH receivers versus the SINR threshold at ID receivers.} \label{fig:SINR}
\vspace{-1em}
\end{figure}

Fig. \ref{fig:SINR} shows the minimum received power at EH receivers versus the SINR threshold $\Gamma$ at ID receivers, where $\Gamma = \Gamma_i, \forall i\in \mathcal {K_I}$, and $P=8$ W are set. It is observed that the proposed design with IRS achieves significant performance gains over the three benchmark schemes. It is also observed that at the medium value of $\Gamma$, the employment of energy beams leads to significant energy enhancement at EH receivers, as compared to the counterpart with information beamforming only. Dedicated energy beamforming is observed to have negligible effects on the performance when $\Gamma$ becomes sufficiently small or large.

\begin{figure}
\centering
 \epsfxsize=1\linewidth
    \includegraphics[width=8cm]{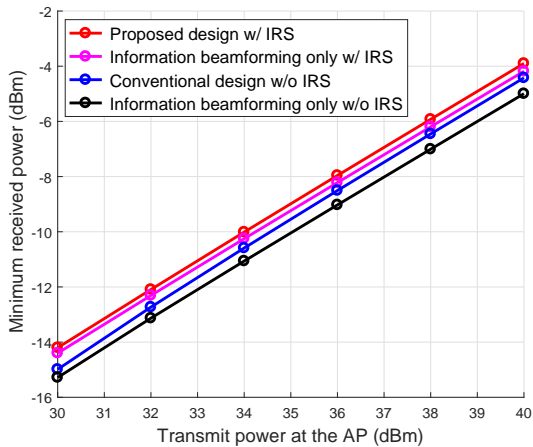}
\caption{The minimum received power versus the transmit power at the AP.} \label{fig:Power}
\vspace{-1em}
\end{figure}

Fig. \ref{fig:Power} shows the minimum received power at EH receivers versus the transmit power $P$ at the AP, where the SINR threshold at ID receivers is set to be $\Gamma_i=15$ dB, $\forall i\in\mathcal{K_I}$. Our proposed design with IRS is observed to outperform the other benchmark schemes over the whole regime of $P$.

\section{Conclusion}
\vspace{-0em}
In this paper, we studied the joint transmit and reflective beamforming design in an IRS-enhanced multiuser SWIPT system with both information and energy signals. Our objective was to maximizing the minimum received power at EH receivers, while ensuring the SINR requirements at ID receivers. We proposed an effective algorithm to solve this non-convex problem by using techniques of alternating optimization and SDR. Numerical results showed that the employment of IRS and dedicated energy beams is crucial to improve both EH and ID performance in such multiuser SWIPT systems.

\end{document}